\documentclass[traditabstract]{aa} 
\usepackage{graphicx}
\usepackage{txfonts}

\def\ltsima{$\; \buildrel < \over \sim \;$}
\def\simlt{\lower.5ex\hbox{\ltsima}}
\def\gtsima{$\; \buildrel > \over \sim \;$}
\def\simgt{\lower.5ex\hbox{\gtsima}}
\def\cgs{{erg cm$^{-2}$ s$^{-1}$}}
\def\ergs{{erg s$^{-1}$}}
\def\cm2{{cm$^{-2}$}}

\def\xrd{{$\chi^{2}_{\rm \nu}$(dof)}}

\def\xnu{{$\chi^{2}(\nu)$}}

\def\fhx{{$F_{\rm 2-10}$}}

\def\lum{{$L_{2-10}$}}
\def\lums{{$L_{0.5-2}$}}

\def\p1{{Paper I}}
\def\fsx{{$F_{\rm 0.5-2}$}}
\def\xmm{{\em XMM--Newton}}
\def\chandra~{{\em Chandra}}

\def\chandra{{\em Chandra}}

\def\xmm{{\em XMM--Newton}}

\def\nh{{N$_{\rm H}$}}
\def\epic{{\em EPIC}}

\def\f14{{10$^{-14}$}}
\def\f13{{10$^{-13}$}}
\def\f12{{10$^{-12}$}}
\def\f11{{10$^{-11}$}}
\def\e22{{10$^{22}$}}

\def\feka{{Fe K$\alpha$}}

\def\eso{{ESO 138$-$G1}}
\def\l58{{$L_{5.8 \mu m}$}}

\begin{document}
   \title{X-ray spectroscopy of the Compton-thick Seyfert 2  ESO 138$-$G1}


   \author{E.~Piconcelli
          \inst{1}, S.~Bianchi\inst{2}, C. Vignali\inst{3}, E. Jim{\'e}nez-Bail{\'o}n\inst{4}, and F. Fiore \inst{1}
         }
 \titlerunning{X-ray observations of ESO 138$-$G1}\authorrunning{E.~Piconcelli et al.}
\offprints{Enrico Piconcelli, \email{enrico.piconcelli@oa-roma.inaf.it}}         
\institute{Osservatorio Astronomico di Roma (INAF), Via Frascati 33, I--00040
  Monteporzio Catone (Roma), Italy  \and Dipartimento di Fisica, Universit\`a degli Studi Roma 3, Via della Vasca Navale 84, I--00146 Roma, Italy \and
Dipartimento di Astronomia, Universit\`a  di Bologna, Via Ranzani 1, I-40127 Bologna, Italy \and 
Instituto de Astronomia, Universidad Nacional Aut\'onoma de M\'exico, Apartado Postal 70-264, 04510 Mexico DF, Mexico
}



 
  \abstract
   {We report on our analysis of \xmm\ observations of the Seyfert 2 galaxy \eso ($z$ = 0.0091). These data reveal a complex spectrum in both its soft and  hard portions. 
The 0.5--2 keV band is characterized by a strong ``soft-excess'' component with several  emission lines, as commonly observed in other narrow-line AGN. Above 3 keV, a power-law fit yields a very flat slope ($\Gamma$ $\sim$ 0.35), along with the presence of a  prominent line-like emission feature around $\sim$6.4 keV. This indicates heavy obscuration along the line of sight to the nucleus. We find an excellent fit to the 3-10 keV continuum with a pure reflection model, which provides strong evidence of a Compton-thick screen, preventing direct detection of the intrinsic nuclear X-ray emission. 
Although a model consisting of a power law transmitted through an absorber with \nh\ $\sim$ 2.5  $\times$ 10$^{23}$ \cm2\ also provides a reasonable fit to the hard X-ray data, the equivalent width (EW) value of $\sim$ 800 eV measured for the \feka\ emission line is inconsistent with a primary continuum obscured by a Compton-thin column density. Furthermore, the ratio of 2-10 keV  to de-reddened [OIII] fluxes for \eso\  agrees with the typical values reported for well-studied Compton-thick Seyfert galaxies. Finally, we also note that  the upper limits to the 15-150 keV flux provided by  {\it Swift}/BAT and {\it INTEGRAL}/IBIS  seem to rule out the presence of a transmitted component of the nuclear continuum even in this very hard X-ray band, hence imply that the column density of the absorber could be as high as 10$^{25}$ \cm2. This makes \eso\ a very interesting, heavy Compton-thick AGN candidate for the next X-ray missions with spectroscopic and imaging capabilities above 10 keV.}
  
 \keywords{Galaxies:~individual:~ESO 138$-$G1 -- Galaxies:~active --
  Galaxies:~nuclei -- X-ray:~galaxies }

   \maketitle
%

\section{Introduction}

Since X-ray radiation is emitted from within a few gravitational radii of the SMBH, it can provide insights into AGN activity with unrivalled detail. Furthermore, typical X-ray observations in the 2-10 keV range easily penetrate regions (i.e. dust lanes, the torus, star-forming knots) with high extinction in optical ($A_V$ $>$ 50 mag) located along our line-of-sight to the nucleus.
The census of AGN in the Universe is, however, still incomplete as illustrated by the mismatch between 
the total mass function of local SMBHs and  the SMBH mass function inferred from the X-ray-selected
AGN luminosity functions (Marconi et al. 2004), 
 and because  only $\sim$ 60\%
of the cosmic X-ray background in the 5-10 keV band has been so far resolved into point sources (Worsley et al. 2005).

It is well-known that one of the  major difficulties to overcome is the detection of the most  elusive AGN, i.e. those suffering heavy obscuration (i.e. with column density of \nh\simgt$\sigma^{-1}_t$ $\approx$1.6 $\times$ 10$^{24}$ \cm2), the so-called Compton-thick (CT) AGN.
As the primary continuum is blocked, their X-ray emission below 10 keV is mainly due to the Compton reflection and fluorescence from optically-thick circumnuclear material (Matt 1997). Accordingly, the observed 2-10 keV fluxes for CT AGN  therefore represent only 1--10\% of their intrinsic fluxes.
The typical  2-10 keV reflection-dominated spectrum  is flat and
characterized by  a prominent (EW \simgt 0.6--1 keV)  Fe K$\alpha$ emission line at 6.4 keV.

According to the X-ray background synthesis models (e.g. Gilli et al. 2007) and direct estimates from local samples of galaxies (Risaliti et al. 1999; Cappi et al. 2006;  Malizia et al. 2009; Burlon et al. 2011), CT AGNs should account for $\sim$ 20--30\% of the entire AGN population.
 Nonetheless, only a few dozen of them (mostly local, low-luminosity Seyfert galaxies) have  been unambiguously discovered and studied so far (Comastri 2004; Fukazawa et al. 2011; Della Ceca et al. 2008 and references therein). Furthermore,  ultra-deep X-ray exposures have  provided clear evidence of powerful CT quasars  at high redshifts (e.g. Feruglio et al. 2011; Comastri et al. 2011; Gilli et al. 2011; Fiore et al. 2011), whose existence has been debated for many years (e.g. Halpern et al. 1999).
Enlarging the sample of well-studied CT AGNs in the X-ray regime is therefore useful  to characterize their properties more accurately, confirm  distinctive spectral features, and 
 strengthen the relations between optical/IR and X-ray observable quantities of this elusive class of X-ray sources.\\

In this paper, we study  the X-ray spectral properties of \eso, a Seyfert 2 galaxy  at $z$ = 0.0091.
This AGN is hosted in an   E/S0 galaxy that has a peculiar morphology of a bulge surrounded by a ring, 
within which  there is a hint of a bar (Alloin et al. 1992; Ferruit et al. 2000). Furthermore, \eso\ exhibits a compact
nucleus  and a bright asymmetric, wedge-shaped 
circumnuclear zone of diffuse light resembling an
ionization cone from the AGN (Munoz-Marin et al. 2007). All these features make \eso\ a  unusual and  particularly interesting object.
On the basis of {\it ASCA} data, Collinge \& Brandt (2000) found a hard spectrum ($\Gamma$ $\approx$ 0.5) and a strong \feka\ emission line, implying that this Seyfert 2 galaxy is  Compton-thick/reflection-dominated. However, statistically acceptable fits can  also be achieved using a partial-covering absorption column of \nh\ = 2 $\times$ 10$^{23}$ \cm2.
Here, we present and discuss the results of our analysis of two archival \xmm\ spectroscopic observations of \eso.

\section{XMM-Newton observations and data reduction}
  \eso\ was observed  by \xmm~(Jansen et al. 2001) on 2007 February 16 (Obs. ID: 0405380201) and
2007 March 25 (Obs. ID: 0405380901).
Both observations were performed with the \epic\ cameras operating in full-frame
mode.

Data were reduced with SAS 11.0 (Gabriel et al. 2004) using standard procedures.
We selected X--ray events corresponding to patterns 0--4(0--12) for the
PN(MOS) camera. The event lists were filtered to
ignore periods of high  background flaring  according to the method
presented in Piconcelli et al. (2004) based on the cumulative
distribution function of background light-curve count-rates.
For the PN detector, source counts were extracted from a circular region with a radius of 30
arcsec (14 arcsec for the 901 observation since it was affected by a higher background level). 
Background counts were estimated from a 50 arcsec-radius region on the same
chip for both datasets.
After screening the final net exposure times are 15.5 and
11.4 ks for the 201 and 901 observation, respectively.
The redistribution matrix files and ancillary response files were
created using the SAS task RMFGEN and ARFGEN, respectively. 
During the \xmm~observations, \eso\ did not display any variation in either  X-ray flux and the spectral shape, thus we coadded  the PN spectra to increase the signal-to-noise ratio and  created a combined response matrix.
Source and background spectra were  summed using the  FTOOL task {\it mathpha}\footnote{Spectral files were combined following  the method outlined in {\it http://heasarc.gsfc.nasa.gov/docs/asca/bgd$\textunderscore$scale/bgd$\textunderscore$scale.html.}}.

The combined PN spectrum was rebinned so that
each energy bin contained at least 25 counts to allow us to use the
$\chi^2$ minimization technique in spectral fitting.

The reduction of MOS spectra was  carried out following the same procedure applied to the PN datasets. We adopted an extraction radius of 31 arcsec for the source spectra, while the background counts were extracted from neighbouring source-free circular regions with a radius of 50 arcsec.

In this paper, we focus on the analysis of  PN spectroscopic data only because of the higher sensitivity of this detector
over the broad 0.5-10 keV range than both MOS cameras (even when they are combined), and above 5 keV in particular. Nevertheless, we checked that consistent results were obtained including the MOS data in our analysis.
Finally,  no useful high resolution RGS data (den Herder et al. 2001) were obtained from the present \xmm\ observations of \eso\ because of their short exposure times.

\section{X-ray spectral analysis}

We analyzed the spectra  with the XSPEC v12.5 package (Arnaud 1996). We limited our analysis to the 0.5--10 keV energy range, where the accuracy of the \epic\ calibrations is maximal. All
models presented in this paper include absorption caused by the line-of-sight Galactic column density of \nh~=
1.3 $\times$ 10$^{21}$ \cm2~(Kalberla et al. 2005). 

We provide best-fit parameter values  in the source frame, unless otherwise specified.  The quoted errors
in the model parameters correspond to a 90\% confidence level for one
interesting parameter ($\Delta\chi^2$ = 2.71; Avni 1976).
In this work, we assume a $\Lambda$CDM cosmology  with $H_{\rm 0}$ = 70 km s$^{-1}$ Mpc$^{-1}$ and $\Omega_{\rm \Lambda}$ = 0.73 (e.g.,  Larson et al. 2011).
\begin{figure*}
\begin{center}
\includegraphics[width=5cm,height=7.5cm,angle=-90]{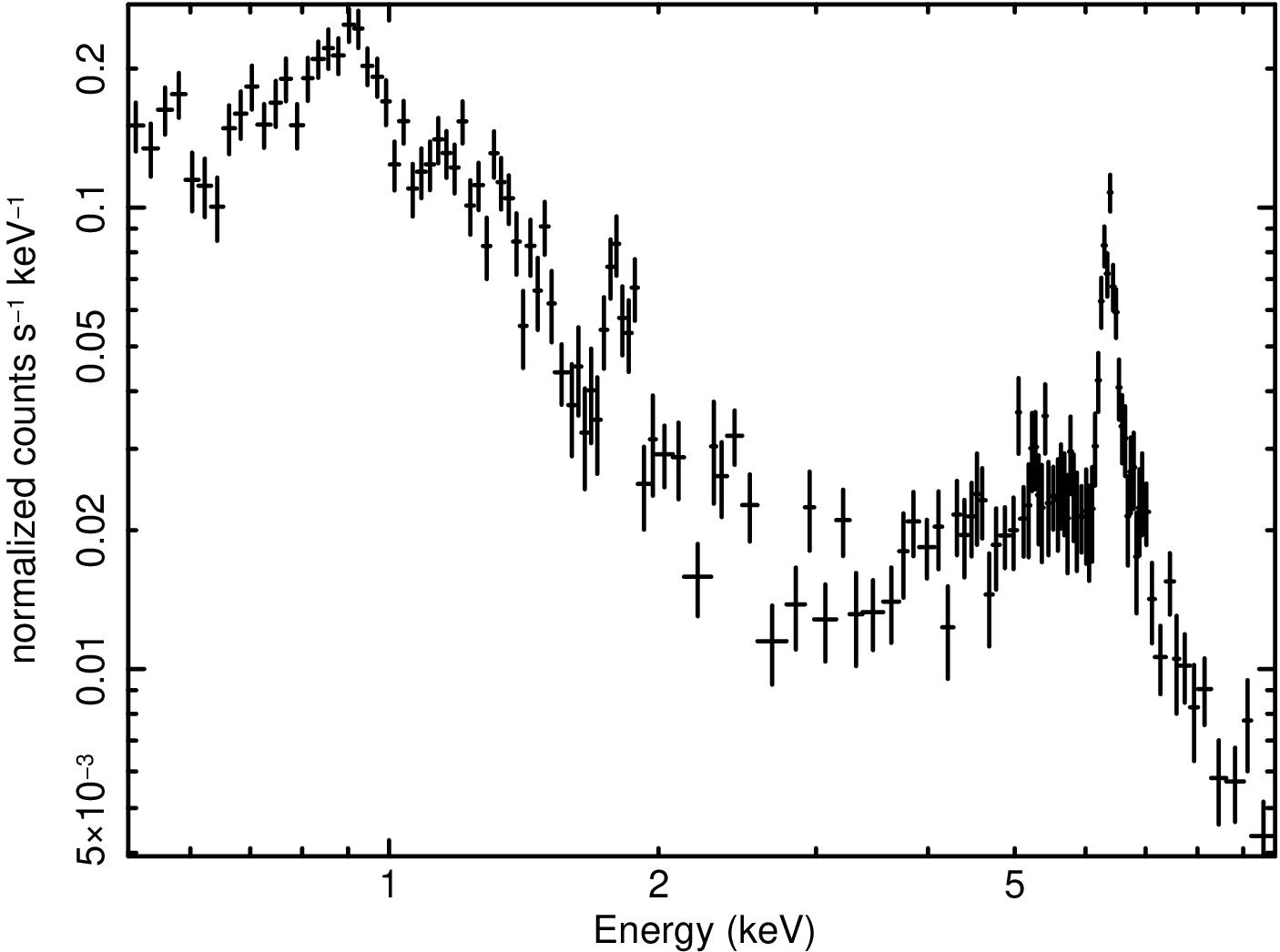}
\hspace{0.5cm}\includegraphics[width=5cm,height=7.5cm,angle=-90]{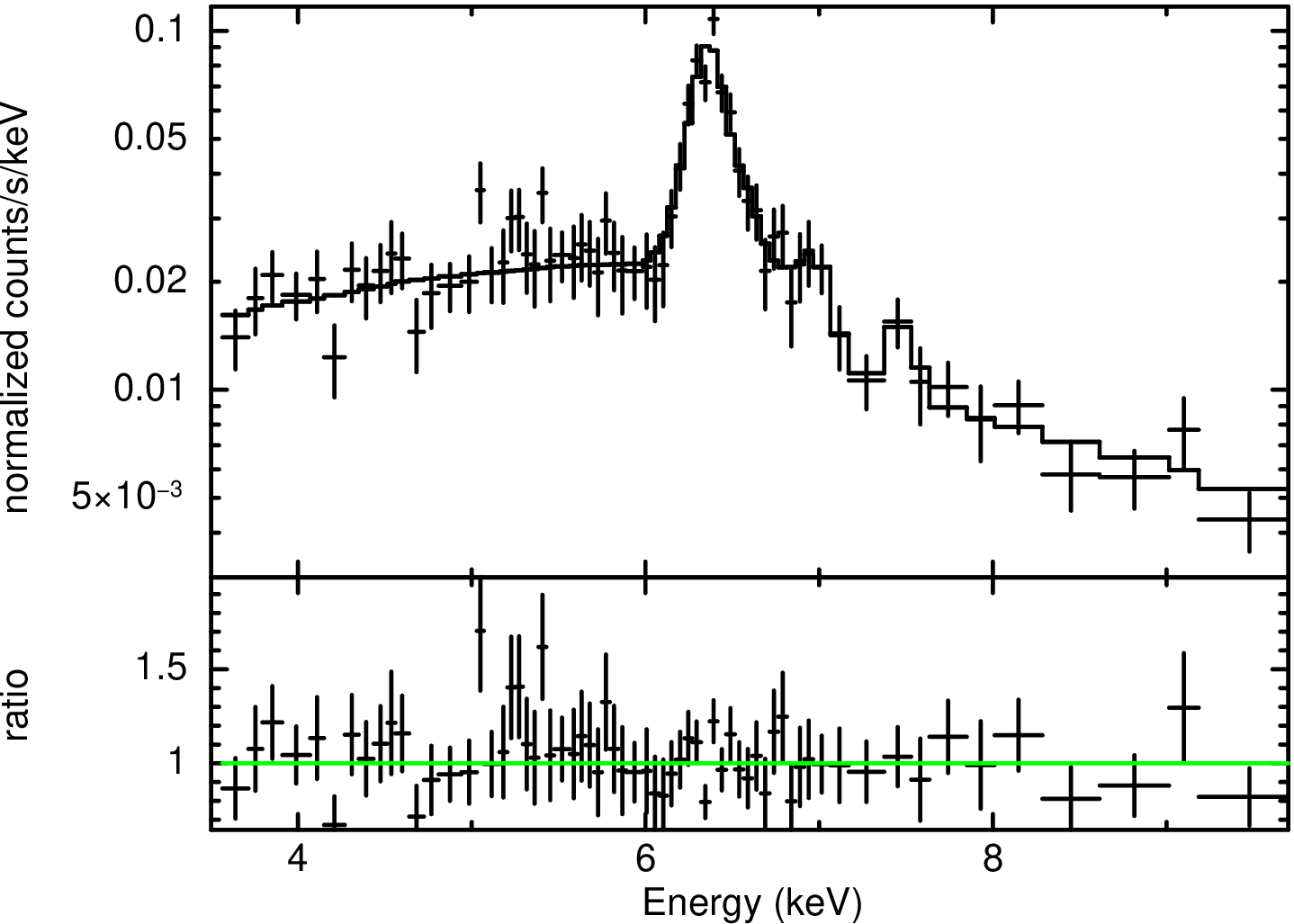}
\caption{(a)--{\it Left:} \xmm\ PN spectrum resulting from the merging of the 2007 February and March observations of \eso.
 (b)--{\it Right:} The RD model fitted to the 3.5-10 keV spectrum of \eso.  This model consists of a Compton reflection continuum component and five Gaussian  lines accounting for the neutral \feka\ line, its Compton shoulder, and the He-like Fe K$\alpha$, Fe K$\beta$, and Ni K$\alpha$ emission lines, respectively.
 The bottom panel shows the data-to-model ratios.}
\label{spettri}
\end{center}
\end{figure*}

\subsection{Hard X-ray band and \feka~lines}
\label{s:spex}
The 0.5--10 keV \epic\ PN spectrum of \eso\ is shown in Fig. 1a. 
Owing to its complexity, we first studied the X-ray spectrum  in the 3.5-10 keV band to
 gain insight into the origin of
 the X-ray emission in this energy range.
 This emission should be relatively unaffected by the extended photoionized/scattered ``soft excess'' component frequently observed in obscured AGNs (Turner et al. 1997; Sambruna et al. 2001; Bianchi et al. 2007), 
which we consider in Sect. 3.2.
As expected for a highly absorbed Seyfert 2 galaxy, a fit to the PN spectrum with a simple power-law in the 3.5-10 keV band (excluding the 6--7.5 keV Fe K energy range) reveals a very flat spectral shape with a photon  index  $\Gamma$ $\sim$ 0.35.
We  tested two alternative scenarios for the hard X-ray spectrum of \eso: (i) a reflection-dominated (RD) one, where the X-ray primary continuum from the AGN is totally depressed in the \epic\ band, as expected in case of  a CT absorber
; and  (ii) a transmission-dominated (TD) scenario, where the AGN emission is absorbed by a Compton-thin (i.e., \nh\ $<$ 10$^{24}$ \cm2) obscuring screen that  only moderately suppresses the X-ray continuum in the 3.5-10 keV range. Moreover, we explore a scenario wherein both reflection and transmission components are present.

In Fig. 1b, we show the hard X-ray PN spectrum fitted by the best-fit RD model, which consists of a cold reflection component, modeled  by {\tt PEXRAV} in XSPEC. In this modelling, we assumed that
 $\Gamma$ = 1.8, that the reflection scaling factor $R$ = $-$1 (defined as $R$ = $\Omega$/2$\pi$, where $\Omega$ is the solid angle subtended by the reflector for isotropic incident emission; Magdziarz \& Zdziarski 1995), that the
metal abundances of the reflector were fixed to their solar values and that the 
inclination angle was fixed to 65 deg. The RD model also included
  a strong fluorescent \feka\ emission line close to 6.4 keV, plus three narrow Gaussian lines at  6.63, 7.058, and 7.54 keV associated with He-like Fe K$\alpha$, Fe K$\beta$, and Ni K$\alpha$ emission, respectively.  The addition of each Fe and Ni K$\alpha$ line produces an improvement in the resulting fit statistic significant at $\geq$99\% confidence level (94\% in the Fe K$\beta$ line case) according to an $F-$test.
 We  fitted the 
Compton  shoulder (CS) to the \feka\ line by adding a Gaussian emission line at 6.3 keV with $\sigma \equiv$40 eV and a normalization fixed to 20\% of the neutral Fe K$\alpha$ line as expected for CT material (e.g., Matt et al. 2002; Yaqoob \& Murphy 2011). The  CS  redwards of  6.4 keV is emitted from neutral material as a result of Compton down-scattering of the Fe line photons in the same medium. The addition of the CS component reduces the $\chi^2$ fit statistic of $\Delta\chi^{2}$ = 6. The RD model gives a very good fit to the data  with a \xnu\ = 78(75) (see Table 1).

The TD model 
yields a reasonable fit to the \xmm~data with a final \xnu\ = 83(73). The spectrum is described by an absorbed  continuum power-law with $\Gamma$ = 1.9$\pm$0.4  and \nh\ = 2.4$^{+0.3}_{-0.4}$ $\times$ 10$^{23}$ \cm2\ (see Table 1).
As for the model RD,  four narrow Gaussian lines (i.e., neutral \feka, He-like Fe  K$\alpha$, Fe K$\beta$, and  Ni K$\alpha$) are included in the fit. When we add a  CS component to the fit that a normalization equal to 10\% of the \feka\ line core (as expected for a column density of $\sim$ 2.5 $\times$ 10$^{23}$ \cm2; Yaqoob \& Murphy 2011), we achieve a $\Delta\chi^{2}$ = 4.

 We investigated whether  a clumpy absorber might be present along our line of sight  by adding an absorbed power-law component to the TD model. The two power laws had the same photon index, thus mimicking a scenario wherein  the obscuring screen along our line of sight is not uniform (e.g., Elitzur 2008; Ramos-Almeida et al. 2009). The fit resulted in \xnu\ = 75(71), with an unusual steep continuum ($\Gamma$ = 2.8$\pm$0.4) seen through two distinct layers of absorption of cold gas with column densities of $\sim$ 7.7 and $\sim$ 1.5  $\times$ 10$^{23}$ \cm2, respectively. However,  the spectral parameters of the dual absorber model were largely unconstrained.  We then fit the data with the continuum slope fixed to $\Gamma$ = 1.9 (as found for the TD model). In this case, the column density of one absorption component was found to be \nh\ $\sim$ 3.5 $\times$ 10$^{23}$ \cm2, while the \nh\ value for the other component was found to be consistent with zero. The ratio of the normalization of two power-law components is $\sim$9.

The hard continuum can also be reproduced with a more complex model where reflection and transmission are both present, which provides a fit with  \xnu\ = 75(72) when  both \nh\ and $\Gamma$ are allowed to vary.
The ratio of the normalization of the reflection to that of the  transmitted component is $\approx$ 6, implying that the reflection continuum can easily account for most of the hard X-ray emission in \eso. However, we note that the quality of our data, along with the lack of spectral coverage above 10 keV, 
does not allow to properly constrain the relative contribution of both components. Fixing the value of the column density of the absorber to \nh\ = 1.6 $\times$ 10$^{24}$ \cm2 ($=$ $\sigma^{-1}_t$) yields a flattening of the photon index to $\Gamma$ $\sim$ 1.85, but does not improve the quality of the fit statistics.

 As performed by Guainazzi et al. (2005) and LaMassa et al. (2011) for their samples of heavily obscured Seyfert 2 galaxies, 
 we  performed a ``local'' fit to the spectrum between 5.5 and 7 keV using  Cash statistics (Cash 1979), which allows the use of unbinned  data without any loss of spectral information,  to obtain  a more robust determination of the neutral \feka\ line parameters.
The continuum  underlying the emission line was modeled with a power law, with the photon index fixed  to the best-fit value of $\Gamma$ = 0.35.
The line energy centroid is at 6.432$\pm$0.015 keV, with an upper limit to the line width of $\sigma$ $<$ 60 eV.
The EW measured with respect to the underlying continuum is  790$\pm$70 eV (see Table 2), including the contribution from the CS (which originates from Compton backscattering of line photons). By fitting the MOS data, the  properties derived for the \feka\ line are fully consistent with those measured by the PN spectrum.

This value of the EW of the 6.4 keV \feka\ emission line unambiguously rules out the TD scenario  (as well as the   Compton-thin dual absorber model) because it is too large for  a continuum obscured by the Compton-thin column density of a few 10$^{23}$ \cm2\ derived from the TD fit to the data (e.g., Ghisellini, Haardt \& Matt  2004; Guainazzi et al. 2005; Fukazawa et al. 2011).
The prominent \feka\ line observed in \eso\ lends strong support to the view that the X-ray primary continuum from the AGN is totally depressed in the \epic\ band, as expected for the presence of a CT absorber, and the 2-10 keV emission
is due to the reprocessing of the AGN light by material surrounding the nucleus.
Consequently, we adopt hereafter the RD model as the best-fit description of the hard X-ray spectrum of \eso.

\begin{table}
\caption{Best-fit spectral parameters for the hard X-ray (3.5-10 keV) continuum  of \eso.
See Sect.~\ref{s:spex} for details.}
\begin{center} 
\begin{tabular}{ccc}
\hline\hline
 \multicolumn{1}{c} {Parameter}&\multicolumn{1}{c} {Model RD$^\dag$}&\multicolumn{1}{c} {Model TD$^\dag$}\\
\hline
 $\Gamma^a$&1.8$^\ddag$&1.9$^{+0.4}_{-0.4}$\\
 Norm$^b$&8.2$^{+0.4}_{-0.4}$&1.6$^{+1.3}_{-0.8}$\\
 N$^{c}_{\rm H}$&$-$&2.4$^{+0.3}_{-0.4}$\\
 $F^{d}_{2-10}$&2.3&2.1\\
 $L^{e}_{2-10}$&4.3 &9.6\\
 \xrd$^{f}$&1.04(75)&1.14(73)\\
\hline
\end{tabular}
\end{center}

$^\dag$ Spectral model: RD ({\it reflection}-dominated), TD ({\it transmission}-dominated); $^a$ photon index of primary continuum; $^b$ normalization of the reflection (RD) or absorbed power-law component (TD) (10$^{-3}$ photons keV$^{-1}$ cm$^{-2}$ s$^{-1}$ at 1 keV); $^c$ column density of the absorber (10$^{23}$ \cm2);  $^d$ 2--10 keV flux  (10$^{-12}$
\cgs); $^e$ 2--10 keV luminosity  (10$^{41}$ \ergs);
 $^f$  reduced $\chi^2$ and number
of degrees of freedom. $^\ddag$  Fixed value (see text for details).
\label{tab:fit}
\end{table}

\begin{table}
\caption{Best-fit spectral parameters for the emission lines detected in the hard X-ray band. See Sect.~\ref{s:spex} for details.}
\label{tab:complex}
\begin{center}
\begin{tabular}{cccc}
\hline\hline
\multicolumn{1}{c} {Line Id.}&
\multicolumn{1}{c} {Energy}&
\multicolumn{1}{c} {EW}&
\multicolumn{1}{c} {Intensity}\\
(1)&(2)&(3)&(4)\\
\hline
\feka$^\ddagger$  &  6.432$^{+0.015}_{-0.015}$  &  790$\pm$70   &2.34$\pm$0.25 $\times$ 10$^{-5}$\\
Fe XXV &  6.63$^\dag$ &  125$^{+54}_{-63}$     &4.5$^{+2.0}_{-2.3}$ $\times$ 10$^{-6}$ \\
Fe K$\beta$ & 7.058$^\dag$  & 80$^{+60}_{-70}$      &2.3$^{+1.7}_{-1.9}$ $\times$ 10$^{-6}$ \\
Ni K$\alpha$& 7.54$^\dag$  &   150$\pm$90   & 2.9$\pm$1.8 $\times$ 10$^{-6}$\\
\hline
\end{tabular} 
\end{center}
The columns give the following information: (1)  line identification; (2) energy (keV); (3) EW (eV); and (4)
 intensity  (photons cm$^{-2}$ s$^{-1}$) of the line. $^\ddagger$ Properties of the \feka\ emission line are derived by a 'local' fit of the unbinned spectrum in the 5.5-7 keV energy range, using the Cash statistics. The EW value includes the contribution of the CS.  $^\dag$  Fixed value.
\end{table}


\subsection{Broad-band spectroscopy}
As shown in Fig. 1a, the soft X-ray portion of the \xmm\ spectrum of \eso\
displays  several  emission features, in addition to  a smooth
excess below $\sim$ 3 keV deviating from the flat shape observed for the hard X-ray band.
As commonly done for low-resolution spectra of heavily obscured AGNs (e.g., Pounds \& Vaughan 2006; Piconcelli et al. 2007), we fitted this ``soft excess''  with a phenomenological model consisting of the RD  model, an unabsorbed power-law component, and a sequence of  narrow Gaussian lines to account for the emission features
observed in the $\sim$ 0.5--2.5 keV range.
  The  power law was found to be steep ($\Gamma_{\rm soft}$ = 3.0$^{+0.5}_{-0.7}$)
and appears to describe a
mixture of emissions caused by the electron-scattered fraction
of the primary continuum, thermal plasma, and  blends of unresolved emission lines and radiative recombination continua, as observed in most X-ray obscured AGNs (Turner et al. 1997; Kinkhabwala et al. 2002). 
Table 3 lists the  best-fit model parameters for the emission lines detected in the soft X-ray band. These emission lines are associated with atomic transitions  commonly detected in the soft X-ray spectra of obscured AGNs.
 However, owing to the limitations imposed by spectral resolution of \epic\ and potential line blending with adjacent emission lines,
both the line identification and parameters should be interpreted with care.

This model provides a good fit to the X-ray spectrum of
\eso\ in the 0.5--10 keV range (Fig. 2), with an associated \xnu\ = 156(133).

 An alternative explanation  of the  ``soft excess'' in some obscured AGNs
is emission related to star-forming activity.
Here we  used a combination of two thermal-emission components ({\tt MEKAL} model in XSPEC) to fit the soft X-ray portion of the spectrum of \eso.
We estimated  a  metallicity of $Z/Z_\odot$ = 0.3$\pm$0.1 and a temperature of $k$T$_1$ $\approx$ 0.7 and $k$T$_2$ $<$ 0.1 keV, respectively.
However, this model yields a poor fit  with \xrd\ = 1.6(152), leaving significant positive residuals in the 1--2.5 keV region and, thus, disfavoring a scenario wherein the soft excess in \eso\ originates in thermal emission from a starburst. 
This is an expected result because a substantial thermal contribution to the soft X-ray emission is typically found only in Seyfert 2/starburst composite galaxies (e.g., Levenson et al. 2001a,b),
while the host galaxy of \eso\ does not exhibit evidence of significant star-forming activity at optical wavelengths (Alloin et al. 1992; Ferruit et al. 2000).

\begin{table}
\caption{Best-fit spectral parameters for the emission lines detected in the soft X-ray band. See Sect. 3.2 for details.}
\label{tab:soft}
\begin{center}
\begin{tabular}{ccc}
\hline\hline
\multicolumn{1}{c} {Energy}&
\multicolumn{1}{c} {Intensity} &
\multicolumn{1}{c} {Identification}\\
(1)&(2)&(3)\\
\hline
0.57$\pm$0.02  & 5.8$^{+2.4}_{-3.0}$  $\times$ 10$^{-5}$& OVII Ly$\alpha$\\
0.72$\pm$0.01  &  3.0$\pm$1.3  $\times$ 10$^{-5}$ & OVII RRC, Fe XVII 3s-2p\\
0.84$\pm$0.02 &  2.5$\pm$0.7  $\times$ 10$^{-5}$ & Fe XVII-XVIII 3d-2p, OVIII RRC\\
0.93$\pm$0.02 &  3.3$\pm$0.9  $\times$ 10$^{-5}$ &NeIX He$\alpha$\\
1.02$^\dag$ &  1.2$^{+0.6}_{-0.9}$  $\times$ 10$^{-5}$    &NeX Ly$\alpha$, Fe XXI 3d-2p\\
1.14$\pm$0.03  & 9.3$\pm$5.8  $\times$ 10$^{-6}$     &Fe XXIII-XXIV L\\
1.21$^{+0.04}_{-0.02}$   & 1.1$\pm$0.5  $\times$ 10$^{-5}$     &NeX Ly$\beta$ \\
 1.34$\pm$0.02  & 1.2$^{+0.2}_{-0.3}$  $\times$ 10$^{-5}$     &MgXI He$\alpha$\\
 1.50$\pm$0.02 & 6.8$^{+1.7}_{-2.0}$  $\times$ 10$^{-6}$     &MgXII Ly$\alpha$\\
 1.82$\pm$0.02 &7.6$^{+1.4}_{-1.9}$  $\times$ 10$^{-6}$      &SXIII He$\alpha$\\
 2.04$^{+0.08}_{-0.06}$ &2.0$^{+1.3}_{-1.6}$  $\times$ 10$^{-6}$      &Si XXIV-XVI Ly$\alpha$ \\
 2.45$^{+0.02}_{-0.03}$ &  6.2$\pm$1.8  $\times$ 10$^{-6}$    &SXV He$\alpha$\\
\hline
\end{tabular} 
\end{center}
The columns give the following information: (1) energy of the line (keV); (2) intensity of the line (photons cm$^{-2}$ s$^{-1}$); (3)
likely identification (e.g., Brinkman et al. 2002; Guainazzi \& Bianchi 2007). $^\dag$ indicates that the energy has been fixed to the laboratory energy of NeX Ly$\alpha$ due to the likely superposition of many unresolved Fe lines.
\end{table}


\section{Fluxes and luminosities}

Once our broad-band best-fit model is assumed, we measure a hard X-ray flux  \fhx\ = 2.3 $\times$ 10$^{-12}$ \cgs\ and a soft X-ray flux \fsx\ = 3.1  $\times$ 10$^{-13}$ \cgs. After correction for Galactic absorption, they correspond to an observed luminosity  \lum\ = 4.3 $\times$ 10$^{41}$ \ergs\ and \lums\ = 8.8 $\times$ 10$^{40}$ \ergs\ in the hard and soft band, respectively.\\

The $T$ ratio of the observed 2-10 keV to the extinction-corrected  [OIII] fluxes can be used as a proxy for the amount of  obscuration of the X-ray primary continuum, since the [OIII] line
is an isotropic indicator of AGN power as it is produced in
the NLR. 
 In particular, X-ray sources for which  $T$ $<$ 1 are typically associated with  a CT absorber and a 2-10 keV RD spectrum (Bassani et al. 1999; Akylas \& Georgantopoulos 2009; Gonzalez-Martin et al. 2009).
The extinction-corrected [OIII] flux of \eso\ is  2.7 $\times$ 10$^{-12}$ \cgs, being derived from the observed value of
  9.72 $\times$ 10$^{-13}$ \cgs\ reported in  Schmitt \& Storchi-Bergmann (1995) after the correction  (e.g., Lamastra et al. 2009 and references therein) for a Balmer decrement H$_\alpha$/H$_\beta$ = 4.25 (Alloin et al. 1992). This gives  $T$ = 0.85, providing further support to the existence of a CT screen that suppresses the observed 2-10 keV flux.  Furthermore,
in the framework of a TD scenario for the hard X-ray spectrum of \eso,  we note that the measured value of $T$ is inconsistent with a Compton-thin  column density of  $\sim$ 2.5 $\times$ 10$^{23}$ \cm2. In this case, the absorption-corrected X-ray flux would be \fhx\ = 5.1  $\times$ 10$^{-12}$ \cgs\ and, therefore, $T$ is $\sim$ 1.9.
However,
as found by many studies (Maiolino et al. 1998; Panessa et al. 2006; Lamastra et al. 2009),  unabsorbed Seyfert 1  and  absorption-corrected Compton-thin Seyfert 2 galaxies indeed typically show  $T$ $\geq$ 10.
\begin{figure*}
\begin{center}
\includegraphics[width=5cm,height=7.5cm,angle=-90]{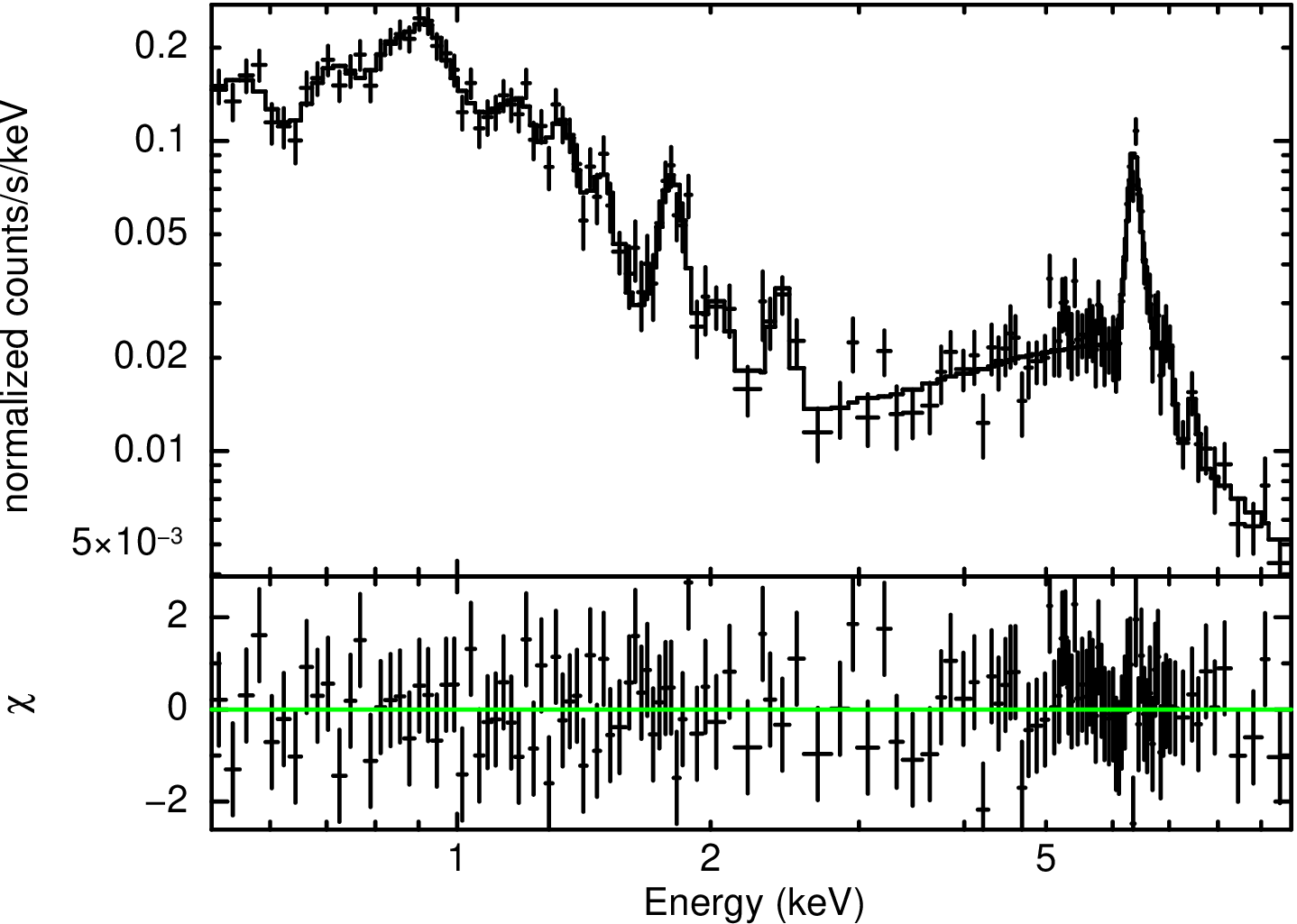}
\hspace{0.5cm}\includegraphics[width=5cm,height=7.5cm,angle=-90]{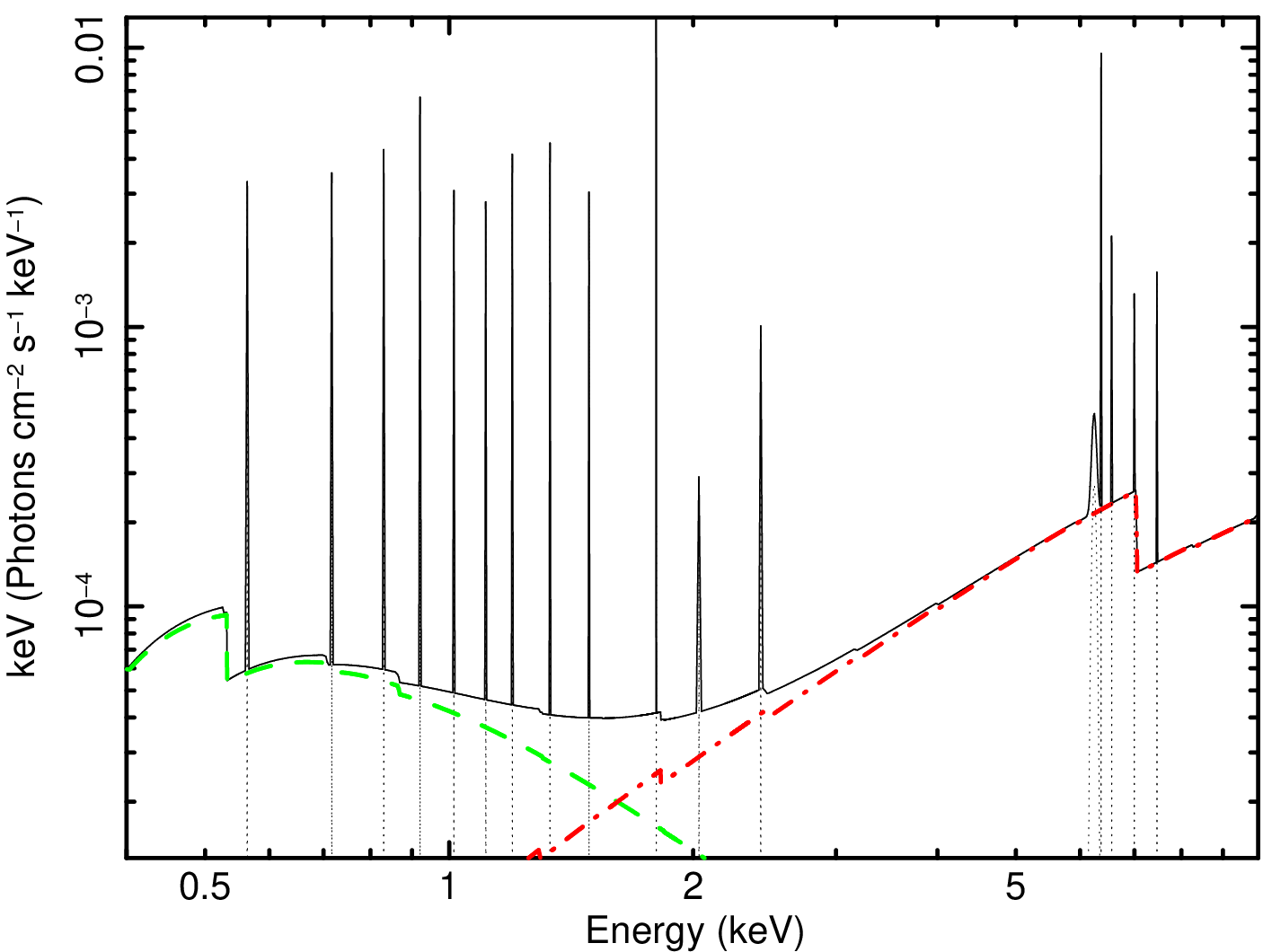}
\caption{(a)--{\it Left:} 
 The 0.5-10 keV spectrum of
  \eso\ when the best-fit model  is applied. 
The lower panel shows the deviations of the observed data from the model in
units of standard deviation. (b)--{\it Right:} Best-fit model for the broad band X-ray spectrum.
This model (solid line) consists of a pure
reflection component (dash-dotted line) plus an additional unobscured power-law component accounting
for the soft X-ray scattered/leaked emission (dashed line). 
It also includes the hard and soft X-ray Gaussian emission lines listed in Tables 2 and 3, respectively.}
\label{broad}
\end{center}
\end{figure*}
         
\section{Discussion}

We have presented the analysis of the combined
spectroscopic data from two  \xmm~observations of the  Seyfert
2 galaxy \eso\ performed in 2007.
The 0.5--10 keV spectrum of this source is typical of a heavily obscured AGN: a very flat  hard X-ray continuum with a strong \feka\ emission line at 6.4 keV and a ``soft excess'' component characterized by  many emission lines from highly ionized metals.
In particular, two main pieces of evidence lead us to consider the RD scenario as the most physically plausible description for the \xmm~data of \eso, thus confirm the  claim by Collinge \& Brandt (2000) based on 1997 {\it ASCA} data for the existence of a CT absorber  covering the primary X-ray continuum in this Seyfert galaxy. 
The first piece of evidence is the value of EW $\sim$ 800 eV for the \feka\ line around 6.4 keV. Such a large value cannot be explained if the underlying continuum is transmitted through a Compton-thin absorber with \nh\ $\sim$ 2.5 $\times$ 10$^{23}$ \cm2, as inferred by the TD fitting model.
A strong \feka~emission line with an EW \simgt~0.7--1 keV is indeed interpreted as
the unambiguous signature of a hard X-ray emission produced by reprocessing  the intrinsic X-ray continuum and then scattering this radiation along the line of sight (Ghisellini, Haardt \& Matt  2004; Guainazzi et al. 2005). 
In this RD scenario, an obscuring screen with \nh\ \simgt 10$^{24}$ \cm2\  is believed to completely block the X-ray primary source from direct view. It has been
generally assumed that the absorber and the reflector consist of the same
material, i.e. the {\it torus}, homogeneously fills a doughnut-shaped region located between the BLR and NLR  (Matt et al. 1996; Molendi et al. 2003). However, the discovery of fast
transitions from a CT to a Compton-thin spectral state (or vice-versa)
have been  observed in a handful of Seyfert galaxies (Risaliti et al. 2005; Bianchi et al. 2009) and interpreted in terms of absorbing  clouds crossing our line of sight  and located very close to the X-ray source (Elitzur 2008). This implies that
  a line-of-sight
clumpy  sub-pc-scale  absorber exists and, most importantly, the coexistence in the same source of multiple absorbing and/or reflecting components distributed across a range of distances from fractions of pc up to several pc from the SMBH.  However, there has so far been no evidence in \eso\ of large variations in the CT absorber. Future X-ray observations of this source might detect RD to TD spectral state transitions, hence reveal both the presence of an inhomogeneous obscuring gas and the intrinsic luminosity of the X-ray continuum. 
Finally, the larger EW value of 1.7$\pm$0.4 keV for the line at 6.4 keV reported by Collinge \& Brandt (2000) can be explained in terms of a blend of the lines associated with Fe K$\alpha$ emission, caused by the poorer spectral resolution of {\it ASCA} detector than that provided by the PN camera.

Secondly, the ratio $T$ = 0.85(1.9) of the observed(intrinsic) 2--10 keV flux to the reddening-corrected [OIII] emission line flux is consistent with a RD X-ray spectrum (e.g., Sect. 4), since this value does not agree with those reported for X-ray sources attenuated by a column density \nh\ $<$ 10$^{24}$ \cm2\ (Maiolino et al. 1988; Bassani et al. 1999; Heckman et al. 2005; De Rosa et al. 2008).
Furthermore, the [OIII] luminosity can be used to infer the intrinsic 2-10 keV luminosity of the AGN in \eso. Using the Eq. (3) in Lamastra et al. (2009), we derive a \lum\ $\approx$ 6  $\times$ 10$^{42}$ \ergs, i.e. well within the typical Seyfert range (and a factor of six larger than the value inferred by assuming the Compton-thin/TD scenario). The ratio of the value of the intrinsic 2--10 keV luminosity estimated by the [OIII] to the  observed value measured from the RD model is $\approx$ 14, which is typical of CT AGN (Comastri 2004; Turner et al. 1997).  A value of \lum\ $\approx$ 10$^{43}$ \ergs\ for \eso\ can also be inferred from its mid-IR luminosity (i.e., $L_{MIR}$ = 3.46 $\times$ 10$^{43}$ \ergs), as shown by Gandhi et al. (2009). We note that the X-ray luminosity calculated with the TD model, i.e. 9.6  $\times$ 10$^{41}$ \ergs\ (see Table 1), is about an order of magnitude lower than this MIR-based estimate.

The estimate of the intrinsic hard X-ray luminosity of \eso\ allows us to derive some information about the SMBH mass of this object.
For a [OIII]-based \lum\  $\approx$ 6  $\times$ 10$^{42}$ \ergs,  Vasudevan \& Fabian (2007) indicate a bolometric correction of 10. Accordingly, we infer an Eddington luminosity $L_{Edd}$ = 6  $\times$ 10$^{44}$ \ergs\ by assuming  a ratio $L/L_{Edd}$ = 0.1, which is typical of Seyfert galaxies in the local Universe (Woo \& Urry 2002). This value of the  Eddington luminosity implies a  black hole mass of 4.6  $\times$ 10$^{6}$ $M_{\odot}$.
It is well-known that near-infrared bulge luminosities and SMBH masses show a tight correlation (e.g., Marconi \& Hunt 2003). The $K$-band magnitude of the bulge of \eso\ is $K$=10.71 mag (Peng et al. 2006), which translates into a $K$-band luminosity of $L_{K, Bulge}$ = 6.6 $\times$ 10$^{10}$ L$_{\odot}$. Accordingly, we find that \eso\ does not lie on the $M_{BH}$-$L_{K, Bulge}$ correlation: the estimated mass of the SMBH is about a factor of ten lower than expected. 
This implies that \eso\ is a peculiar  early-type galaxy with an under-massive black hole,
or an intrinsic X-ray luminosity ten times higher than that estimated based on the [OIII] luminosity, thus  providing further support for a  CT nature of this AGN.\\ 

Finally, useful insights into the nature of the obscuring screen along the line of sight to the nucleus in \eso\ can also be derived from the 15-150 keV  flux level constraint.
We note that \eso\ is indeed included in the catalogs of hard X-ray sources detected by {\it INTEGRAL} IBIS (Bird et al. 2010) and {\it Swift} BAT  (Cusumano et al. 2010) above 15 keV. Specifically,
it is associated with the BAT source 2PBC J1651.9$-$5914, for which $F_{15-150}$ = 2.3$\pm$0.5  $\times$ 10$^{-11}$ \cgs,
 that is centered at a  distance of $\sim$ 3 arcmin from \eso.
On the basis of the RD model (i.e. assuming a broad-band, pure reflection spectrum), it is possible to estimate\footnote{We note, however, that any accurate estimate of the reflected emission  depends on the geometry of the reflection medium (e.g., Ikeda et al. 2009), which is basically unknown. Meaningful constraints on the geometry of the reprocessor will be possible only using the high signal-to-noise ratio,  broad-band spectral data expected from observations performed  by {\it ASTRO-H} and {\it NuStar} in the near future.} a 15-150 keV flux of 3.1$^{+0.1}_{-0.2}$ $\times$ 10$^{-11}$ \cgs.

However, since the angular resolution of BAT is $\sim$ 20 arcmin FWHM and the X-ray loud starburst/Seyfert 2 composite galaxy NGC 6221 (e.g. Levenson et al. 2001a) is located at 11 arcmin from our target, the BAT flux can  only be used  as an upper limit to the X-ray emission from \eso\ in the 15-150 keV band (V. La Parola, private communications).  The same argument is also valid for the {\it INTEGRAL} IBIS flux reported in Bird et al. (2010).
We were also able to extract the PN spectrum of NGC 6221, as it falls within the field of view of the detector (although close to the edge) in our observations, to help us determine  the possible contribution of this source to the observed BAT flux.
The X-ray spectrum of NGC 6221 is well-fitted by a combination of a soft thermal plasma component ($k$T $\sim$ 0.6 keV)
and  a Compton-thin (\nh\ $\sim$  10$^{21}$ \cm2) absorbed power law with $\Gamma$ = 1.6$\pm$0.1, as found by Levenson et al. (2001a) based on {\it ASCA} 
data\footnote{The 2-10 keV flux of NGC 6221 measured by the \xmm\ observation is \fhx\ $\sim$ 3 $\times$ 10$^{-12}$ \cgs, i.e. a factor of 4.5 lower than the value derived by previous {\it ASCA} data, thus confirming the presence of large flux variability in this source as reported in Levenson et al. (2001b).}.
By extrapolating this model to energies $>$ 10 keV, we estimate a value of $F_{15-150}$ = 1.1 $\times$ 10$^{-11}$ \cgs\
for the flux in the 15--150 keV band.
 Therefore, a sizable fraction of the BAT  flux ascribed to \eso\ in the Cusumano et al. catalog might be due to the emission from NGC 6221: in particular,  it is likely that  the total 15-150 keV flux of $F_{15-150}$ $\approx$ 2.3 $\times$ 10$^{-11}$ \cgs\ seen by BAT and IBIS is produced roughly equally by  the two sources.
In turn, this may indicate that the column density of the nuclear absorber in \eso\ should be \nh \simgt\ 10$^{25}$ \cm2, otherwise any additional emission component in the 15-150 keV band caused by the X-ray primary continuum transmitted through an absorber with \nh\ $\sim$ a few  10$^{24}$ \cm2, would produce a further increase in the final flux, significantly exceeding the IBIS/BAT flux values.

In this respect, future observations of \eso\ above 10 keV will be crucial to shed light on the properties of the X-ray continuum emission and the obscuring gas in the nuclear
environment of this Seyfert 2 galaxy. We note, however, that
given the proximity of NGC 6221, low-resolution (i.e. with FWHM $\gg$ 10 arcmin) observations carried out with current X-ray observatories such as {\it Swift} BAT,    {\it INTEGRAL} IBIS and {\it Suzaku} PIN cannot provide an unambiguous and definitive test  of whether \eso\ contains a heavily CT (i.e. \nh\ $\gg$ 10$^{24}$ \cm2) absorber. This will be achieved by
next generation X-ray missions  such as {\it NuSTAR} (Harrison et al. 2010) and {\it ASTRO-H} (Takahashi et al. 2010), which should  to fly in 2012 and 2014, respectively, as they combine $\sim$ 50--90 arcsec imaging with high spectral resolution in the $\sim$ 10--60 keV energy range.\\

The soft  portion of the X-ray spectrum of \eso\ shows the typical properties of other obscured AGNs, with the presence of a highly structured ``soft excess''. High-resolution imaging and spectroscopy of  Seyfert 2 galaxies performed with \xmm\ and \chandra\
have allowed us to reveal that the soft X-ray spectrum originates from extended emission photoionized by the AGN (Kinkhabwala et al. 2002; Bianchi et al. 2006; Guainazzi \& Bianchi 2007). The contribution of collisionally ionized plasma associated with
 starburst regions does not appear to be significant, except in objects known to host intense star-formation activity, such as highly disturbed systems and very luminous IR galaxies (Netzer et al. 2005; Guainazzi et al. 2009; Piconcelli et al. 2010).
Narrow-band [OIII] images of \eso\ show a jet-like feature extending up to $\approx$ 2 kpc westwards from the center, produced by the ionizing nuclear radiation  (Schmitt \& Storchi-Bergmann 1995).
Furthermore, high-resolution near-UV and [OIII] images of \eso\  taken with the ACS and WFPC2 detectors onboard {\it HST}  (Munoz-Marin et al. 2007; Ferruit et al. 2000) have revealed a bright,  ionization cone-shaped
circumnuclear zone ($\sim$ 200 pc)  of diffuse light. 
Close morphological correspondences between the soft X-ray and the [OIII] emission on a kpc scale are frequently observed in heavily obscured Seyfert 2 galaxy and clearly indicate a common physical origin (i.e. photoionization) for the two emissions (Brinkman et al. 2002; Bianchi et al. 2006 and references therein).
Accordingly, we suggest that this may turn out to also be the case  for the origin of the line-rich ``soft excess'' observed in the \xmm\ spectrum of \eso.
However,  high-resolution X-ray spectroscopy and  imaging with deep exposures of \eso\  are needed  to unambiguously confirm the presence of photoionized gas in this Seyfert 2 galaxy and resolve its morphological details.

\begin{acknowledgements}
We thank the referee for comments that helped to improve the clarity of this manuscript.
We would like to thank the staff of the \xmm~Science Operations Center for
their support. We thank Valentina La Parola for her help in the discussion about {\it Swift} BAT observations.
E.P., S. B., and C.V. acknowledge support under ASI/INAF contracts I/088/06/0 and  I/009/10/0.
Based on observations obtained with \xmm, an ESA science mission with instruments and contributions directly funded by ESA Member States and NASA.
This research has made use of the NASA$/$IPAC
Extragalactic Database (NED) which is operated by the Jet Propulsion
Laboratory,  California Institute of Technology, under contract with
the National Aeronautics and Space Administration.
\end{acknowledgements}
{}
\end{document}